\documentclass[a4paper,twoside]{article}

\usepackage{epsfig}
\usepackage{calc}
\usepackage{amssymb,amsfonts}
\usepackage{graphicx}
\usepackage{textcomp}
\usepackage{xcolor}
\usepackage{comment}

\usepackage{subfig}

\newcommand{\beq}{\begin{equation}}
\newcommand{\eeq}{\end{equation}}
\usepackage{amstext}
\usepackage{amsmath}
\usepackage{amsthm}
\usepackage{multicol}
\usepackage{pslatex}
\usepackage{apalike}
\usepackage{algorithm2e}
\usepackage[bottom]{footmisc}
\usepackage{SCITEPRESS}     

\begin{document}

\title{Temporal Complexity of a Hopfield-Type Neural Model in Random and Scale-Free Graphs}

\author{\authorname{Marco Cafiso \sup{1},\sup{2} and Paolo Paradisi\sup{2},\sup{3}}
\affiliation{\sup{1}Department of Physics 'E. Fermi', University of Pisa, Largo Bruno Pontecorvo 3, I-56127, Pisa, Italy}
\affiliation{\sup{2}Institute of Information Science and Technologies ‘A. Faedo’, ISTI-CNR, Via G. Moruzzi 1, I-56124, Pisa, Italy}
\affiliation{\sup{3}BCAM-Basque Center for Applied Mathematics,\\ Alameda de Mazarredo 14, E-48009, Bilbao, BASQUE COUNTRY, Spain}
\email{marco.cafiso@phd.unipi.it, paolo.paradisi@cnr.it}
}


\abstract{The Hopfield network model and its generalizations were introduced
as a model of associative, or content-addressable, memory. They
were widely investigated both as an unsupervised learning method in 
artificial intelligence and as a model of biological neural
dynamics in computational neuroscience.
The complexity features of biological neural networks have attracted the scientific community's interest for the last two decades. More recently, concepts and tools borrowed from complex network theory were applied to artificial neural networks and learning, thus focusing on the topological aspects.
However, the temporal structure is also a crucial property displayed by biological neural networks and investigated in the framework of systems displaying complex intermittency.
The Intermittency-Driven Complexity (IDC) approach indeed focuses on the 
metastability of self-organized states, whose signature is a power-decay in the inter-event time distribution or a scaling behaviour in the related event-driven diffusion processes.
The investigation of IDC in neural dynamics and its relationship with network topology is still in its early stages.
In this work, we present the preliminary results of an IDC analysis carried out on a bio-inspired Hopfield-type neural network comparing two different connectivities, i.e., scale-free vs. random network topology. 
We found that random networks can trigger complexity features similar to that of scale-free networks, even if with some differences and for different parameter values, in particular for different noise levels.}

\maketitle

\section{\uppercase{Introduction}}
\label{sec:intro}

The Hopfield model is the first example of a recurrent neural network defined by a set of linked two-state McCulloch-Pitts neurons evolving in discrete time.
The Hopfield model is similar to the Ising model \cite{Ising1925beitrag} describing the dynamics of a spin system in a magnetic field, but with all-to-all connectivity among neurons instead of local spin-spin interactions.
More importantly, in his milestone paper \cite{hopfield_pnas1982}, Hopfield first introduced a rule for changing the topology of network connections based on external stimuli.
Hopfield first proposed and investigated the properties of this network model in \cite{hopfield_pnas1982} and in successive works \cite{hopfield_pnas1984,hopfield_n1995}. In particular, he also proposed an extension of the original 1982 model to a continuous-time leaky-integrate-and-fire neuron model \cite{hopfield_pnas1984}, also considering the case of neurons with graded response,
i.e., with a sigmoid function mediating the voltage inputs from upstream neurons.
The main property of the Hopfield model is that the connectivity matrix is allowed to change according to a Hebbian rule \cite{hebb_1949}, which is often summarised in the statement:
``(neuron) cells that fire together wire together'' \cite{lowel_s1992}. 
To our knowledge, with this rule, the Hopfield neural network model results in being the first model used for the investigation of associative, or content-addressable, memory.
In the Artificial Intelligence (AI) jargon, external stimuli correspond to the cases of a training dataset that trigger the changes in the connectivity matrix. This process involves decoding the input data into a map of neural states (see, e.g., \cite{hopfield_n1995}).

The Hopfield neural model, as such and its variations, belong to the class of Spiking Neural Networks (SNNs). SSNs are still attracting considerable attention due to energy efficiency and high sensitivity to temporal features of data and, even if they are nowadays less efficient concerning the classical deep neural networks, are thought to have great potential in the context of neuromorphic computing \cite{davies_ieeem2018}.

\noindent
An interesting aspect studied by Hopfield is the emergence of collective,
i.e., self-organizing behaviour in relation to the stability of memories in the network model.
In this framework, Grinstein et al. \cite{grinstein2005model} investigated the role of
topology in a neural
network model extending the Hopfield model to a more biologically plausible one, but partially maintaining the computational advantage of two-state McCulloch-Pitts neurons with respect to continuous-time extension of the model.
This was achieved by introducing a maximum firing time and a refractory time in the single neuron dynamics.

\noindent
Since the last two decades, the interest towards the complex topological features of neural networks has gained momentum in many scientific fields involving concepts and tools of computational neuroscience and/or AI (see, e.g., \cite{kaviani_eswa2021} for a survey). In particular, neural networks with complex topologies, such as random (Erd$\ddot o$s-R\' enyi) \cite{erdos1959random}, \cite{gros_2013}, small-world or scale-free networks \cite{boccaletti_pr2006}, were shown to outperform artificial neural networks with all-to-all connectivity \cite{mcgraw_pre2003,torres_nc2004,lu_pla2006,shafiee_ieee-a2016,kaviani_icte2020,adjodah_proceed2020,kaviani_eswa2021}.

Complexity is a general concept related to the ability of a multi-component system
to trigger self-organizing behaviour, a property that is manifested in the generation of spatio-temporal coherent states
\cite{paradisi_csf15_preface,grigolini_csf15_bio_temp_complex,paradisi_springer2017}.
Interestingly, in many research fields, many authors consider the complexity of a system as a concept essentially referring to its topological structure, which 
is an approach borrowed from graph theory and complex networks \cite{watts_n1998,barabasi_s1999,albert_rmp2002,barabasi_nrg2004}.
However, another aspect, which is often overlooked and instead is typically
a crucial feature of complex self-organizing behaviour, is the temporal structure of the system. This is intimately related not only to
the topological/geometrical structure of the network but also to its dynamical properties, both at the level of single nodes, of clusters of nodes, and as a whole.
Hereafter we refer to Temporal Complexity (TC), or Intermittency-Driven Complexity (IDC), as the property of the system to generate metastable self-organized states, the duration of which is marked by rapid transition events between two states \cite{grigolini_csf15_bio_temp_complex}. 
The underlying theories of TC/IDC refer to
Cox's renewal theory \cite{cox_1970_renewal}, Cox's failure events are reinterpreted in a temporal sense, that is, precisely as rapid transitions or jumps in the system's observed variables.
The rapid
transitions can occur between two self-organized states or between a self-organized state and a disordered or non-coherent state. The sequence of transition events is
then described as a point process and the ideal condition for TC/IDC is  the renewal
one \cite{cox_1970_renewal,bianco_cpl07,paradisi_cejp09}, which is not easily determined being mixed to
spurious effects such as secondary events and noise 
\cite{paradisi_springer2017,paradisi_csf15_pandora}.
The self-organizing behaviour can be detected by the recognition of given patterns
in the system's variables, e.g., eddies in a turbulent flow or synchronization epochs
in neural dynamics, and the identification of events is achieved by means of 
proper event detection algorithm in signal processing \cite{paradisi_springer2017,paradisi_chapter2023}.

\noindent
A general concept commonly accepted in the complex system research field is that
complexity features are related to the power-law behaviour of some observed features, e.g.:
space and/or time correlation functions, the distribution of some variables such as
the sequence of inter-event times and the size of neural avalanches 
\cite{beggsplenz_jn2003}.
In TC/IDC a crucial feature to be evaluated is the probability density function (PDF) of 
inter-event times, or Waiting Times (WT). However, the WT-PDF is often blurred by
secondary events related to noise or other side effects \cite{allegrini_pre10,paradisi_csf15_pandora}. A more reliable analysis relies on
diffusion processes derived by the sequence of events and on their scaling analysis \cite{akin_jsmte09}.

This approach involves several scaling analyses widely investigated in the
literature that were integrated in the Event-Driven
Diffusion Scaling (EDDiS) algorithm \cite{paradisi_springer2017,paradisi_csf15_pandora}.
This approach was also successfully applied in the
context of brain data, being able to characterize
different brain states from wake, relaxed condition to the different sleep stages \cite{paradisi_aipcp13,allegrini_csf13,allegrini_pre15}.

\noindent
At present, the relationships between network connectivity and temporal complexity by one hand and the complexity of connectivity matrix and learning efficiency are still not clear.
In this work, we present some preliminary results regarding the first aspect, having in
mind the potential applications regarding the 
second aspect, i.e., connectivity vs. learning efficiency.
Along this line, an IDC analysis is carried out on a bio-inspired Hopfield-type neural network comparing two different connectivities, i.e., scale-free vs. random network topology. 
In Section \ref{sec:model} we introduce the methods to generate the network topologies and the details of the bio-inspired Hopfield-type neural network model. Section \ref{sec:eddis} briefly describes the event-based scaling analyses. In Section \ref{sec:results} we describe the results of numerical simulations and of their IDC analyses that are discussed in Section \ref{sec:discussion}.
Finally, we sketch some conclusions in Section \ref{sec:conclusion}.

\section{\uppercase{Model description}}
\label{sec:model}

\subsection{Network topology: scale-free vs. Erd$\ddot o$s-R\'enyi}

We here consider two types of networks, both with $N$ number of nodes\footnote[1]{
Let us recall that the degree of a node in the network is the number of links 
of the node itself. In a directed network, each node has an out-degree, given
by the number of outgoing links, and an in-degree, given by the number of ingoing links.
}.
Our networks are
constrained to have the same minimum number $k_0$ of links for each neuron and the same average number of links $\langle k \rangle$. In both cases, self-loops and multiple directed edges from one node to another are excluded, following the methodology outlined in \cite{grinstein2005model}. 
The first class of networks is that of Scale-Free (SF) graphs, characterized by a power-law node out-degree distribution. 
The probability of a node \(i\) to
have \(k_i\) outgoing links is given by: 
\begin{eqnarray}
&&\forall i = 1, ..., N:\ \ P_{_{\rm SF}}(k_i) = \frac{m}{k_i^{\alpha}} \\
\label{sf_1}
\ \nonumber \\
&&m = \frac{ \alpha-1 }{ k_0^{(1-\alpha)} - (N-1)^{(1-\alpha)} } 
\label{sf_2}
\end{eqnarray}
For the construction of SF networks, we used a power-law exponent \(\alpha = 2.5\).

\noindent
The second class of networks is that of Erd$\ddot o$s-R\'enyi (ER) graphs, which are random graphs where each pair of distinct nodes is connected with a probability \(p\). In an ER network with $N$ nodes and without self-loops, the average degree is simply
given by: $\langle k \rangle_{_{\rm ER}} = p_{_{\rm ER}} \left(N-1\right)$. Then, from the
equality of degree averages: $\langle k \rangle_{_{\rm ER}} = \langle k \rangle_{_{\rm SF}}$
we simply derive:
\beq
p_{_{\rm ER}} = \frac{\langle k \rangle_{_{\rm SF}}}{N-1}
\label{p_random}
\eeq
The theoretical mean out-degree of the SF network is approximated by the following formula:
$$
\langle k \rangle \simeq m \frac{k_0^{2-\alpha} - (N-1)^{2-\alpha}}{\alpha-2} = $$
$$
= \frac{\alpha-1}{\alpha-2} 
\frac{k_0^{2-\alpha} - (N-1)^{2-\alpha}}{k_0^{1-\alpha} - (N-1)^{1-\alpha}}
$$
that is obtained by considering $k$ as a continuous random variable.
However, due to the large variability of SF degree distribution, different 
statistical samples drawn from $P_{_{\rm SF}}$ can have very different mean outdegrees.
Thus, we have chosen to numerically evaluate the mean out-degree associated with the
sample drawn from $P_{_{\rm SF}}$ and to use this value instead of the theoretical one
to define $P_{_{\rm SF}}$.

\noindent
The algorithm used to generate the two network topologies is as follows:
\begin{enumerate}
    \item[(SF)]
    \begin{enumerate}       
        \item
        For each node \(i\), choose the out-degree \(k_i\) 
        as the nearest integer of the real number defined by:    
        \beq
            \label{SF_out_degree}
            k_i = (((N-1)^{(1-\alpha)} - k_0^{(1-\alpha)})\xi_i + k_0^{(1-\alpha)})^{\frac{1}{1 - \alpha}}
        \eeq
        being $\xi$ a random number uniformly distributed in $[0,1]$. This formula is obtained by the cumulative function method. The drawn $k_i$ are within the range $[k_0 , N-1]$.\\
        \item 
        Given $k_i$ for each node \(i\), the target nodes are selected by drawing $k_i$ integer numbers $\{j^i_1,...,j^i_{k_i} \}$ uniformly distributed in the set 
        $\{1,...,i-1,i+1,...,N \}$.      
        \item 
        Finally, the adjacency or connectivity matrix is defined as: 
        \beq
        \label{adjacency_sf}
           A^{^{SF}}_{ij} = \left\{
           \begin{array}{ll} & 1 \quad {\rm if}\ \  
           j\in T^i = \{ j^i_1,...,j^i_{k_i} \}\\
           & 0 \quad {\rm otherwise}
           \end{array}
           \right.
        \eeq
     With this choice, the in-degree distribution results in a mono-modal distribution similar to a Gaussian distribution.        
\end{enumerate}
 \item[(ER)]
\begin{enumerate}
\item
From the adjacency matrix $A^{^{SF}}_{ij}$ the actual mean out-degree $\langle k \rangle_{_{\rm SF}}$ is computed.
\item
For each couple of nodes $(i,j)$ with $j \ne i$ a random number $\xi_{i,j}$ is drawn from a uniform distribution in $[0,1]$.
\item 
Finally, the adjacency matrix is defined as: 
  \beq
        \label{adjacency_er}
           A^{^{ER}}_{ij} = \left\{
           \begin{array}{ll} & 1 \quad {\rm if}\ \ 
            \xi_{i,j} < p_{_{\rm ER}}  \\
           & 0 \quad {\rm otherwise}
           \end{array}
           \right.
        \eeq
where $p_{_{\rm ER}}$ is given by Eq. \eqref{p_random}.
\end{enumerate}
\end{enumerate}

\subsection{The Grinstein Hopfield-type network model}

Grinstein et al. \cite{grinstein2005model} modify the Hopfield network by adding three elements: (\(i\)) a random endogenous probability of firing \(p_{endo}\) for each node; (\(ii\)) a maximum firing duration, thanks to which the activity of a node shuts down after \(t_{max}\) time steps; (\(iii\)) a refractory period such that a node, once activated and subsequently deactivated, must remain inactive for at least \(t_{ref}\) consecutive time-steps. 
Each neuron $i$ has two states: \(S_i = 0\) ("not firing") and \(S_i = 1\) ("firing at maximum rate"). The weight of link from $j$ to $i$ is
given by \(J_{ij}\) (Nonconnected neurons have \(J_{ij} = 0\)). 
The network is initialized at time \(t = 0\) by randomly setting each neuron state \(S_i(0)\) equal to 1 with a probability \(p_{init}\) that we chose equal to the endogenous firing probability. At each time step the weighted input to node \(i\) is defined:
\begin{equation} \label{grinstein_weigthed_input_to_node_j}
    I_i(t) = \sum_j J_{ij}S_j(t)
\end{equation}
as in the Hopfield network dynamics. The state of the node \(i\) evolves with \(t\) according to the following algorithm:
\begin{enumerate}
    \item If \(S_i(\tau)\) for all \(\tau = t, t-1,\cdots,t-t_{max}+1\), then \(S_i(t+1) = 0\) (maximum firing duration rule).
    \item If \(S_i(t) = 1\) and \(S_i(t+1) = 0\) then $S(\tau)=0$ for 
    \( (t + 1) < \tau \leq (t + t_{ref}) \)  
  (refractory period rule).
    \item If neither rule 1 nor rule 2 applies, then
    \begin{enumerate}
        \item If \(I_i(t) \geq b_i\) then \(S_i(t + 1) = 1\);
        \item If \(I_i(t) < b_i\) then \(S_i(t + 1) = 1\) with a probability equal to \(p_{endo}\) otherwise \( S_i(t + 1) = 0 \).
    \end{enumerate}
    where $b_i$ is the firing threshold of neuron $i$.
\end{enumerate}
In our study, unlike Grinstein and colleagues, the link's weights and the firing thresholds are taken uniformly throughout the network:  \(J_{ij} = J\) and \( b_i = b \).

\section{\uppercase{Event-driven diffusion scaling Analysis}}
\label{sec:eddis}

The diffusion scaling analysis is a powerful method for scaling detection and, when
applied to a sequence of transition events, can give useful information on the underlying dynamics that indeed generate the events.
The complete IDC analysis involves the Event-Driven Diffusion Scaling (EDDiS) algorithm 
\cite{paradisi_csf15_pandora,paradisi_springer2017} 
with the
computation of three different random walks generated by applying three walking rules
to the sequence of observed transition events and the computation of the second moment scaling and the similarity of the diffusion PDF. 
The general idea is based on the Continuous Time Random Walk (CTRW) model \cite{montroll1964random,weiss1983random}, where a particle moves move only at the event occurrence times.
Here we limit to the so-called Asymmetric Jump (AJ) walking rule \cite{grigolini2001asymmetric}, which simply consists of making a unitary jump ahead when an event occurs and, thus, corresponds to the counting process generated by the event sequence: 
\beq
X(t) = \#\{n: t_n < t\}.
\label{ctrw_aj}
\eeq
The method used to extract events from the simulated data and the
scaling analyses are described in the following.

\subsection{Neural coincidence events}

The IDC features here investigated are applied to {\it coincidence events} that are defined as the events at which a minimum number $N_c$ of neuron fires at the same time.
Then, given the global set of single neuron firing times, the coincidence event time is defined as the occurrence time of more than \(N_c\) nodes firing simultaneously, i.e.,
in a tolerance time interval of duration \(\Delta t_c\). 
Hereafter we always set \(\Delta t_c\) equal to a sampling time, i.e., 
\( \Delta t_c = 1\), which is equivalent to look for 
simultaneous events.  
The total activity distribution of the network corresponds to the size distribution of 
coincidences with minimum number $N_c=1$: $P(n_c| N_c=1)$. The actual threshold
$N_c$ here applied is defined by computing the $35$th percentile of 
$P(n_c| N_c=1)$.
Then, each $n$-th coincidence event is described by its occurrence time $t_c(n)$ and its size $S_c(n)$.

\subsection{Detrended Fluctuation Analysis (DFA)}

\noindent
DFA is a well-known algorithm (see, e.g., \cite{peng_pre94}) that is widely used in the literature for
the evaluation of  the second-moment scaling $H$ defined by:
\begin{eqnarray}
    &F^{^2}(\Delta t) = \langle \left( \Delta X(\Delta t) - \Delta X_{\rm trend}(\Delta t) \right )^2 \rangle \sim t^{2H}
    \label{h_scaling}\\
    \ \nonumber \\
    & F(\Delta t) = a \cdot \Delta t^H  \Rightarrow \nonumber \\
    & \Rightarrow  \log(F(\Delta t)) = \log(a)+ H \cdot \log(\Delta t) 
    \label{dfa_function}
\end{eqnarray}
being $\Delta X (t,\Delta t) = X(t+\Delta t) - X(t)$.
We use the notation $H$ as this scaling exponent is essentially the same as the classical Hurst similarity exponent \cite{hurst_1951}.
$X_{\rm trend}(\Delta t)$ is a proper local trend of the time series.
The DFA is computed over different values of the time lag $\Delta t$
and the statistical average is carried out over a set of time windows
of duration $\Delta t$ into which the time series is divided.
In the EDDiS approach, DFA is applied to different event-driven diffusion
processes (see, e.g., \cite{paradisi_npg12,paradisi_springer2017}).
To check the validity of Eq. \eqref{dfa_function} on the data and evaluate
the exponent $H$ is it sufficient to carry out a linear best fit in
the logarithmic scale.
To perform the DFA we employed the function MFDFA of package MFDFA of Python \cite{gorjao2022MFDFA}.

\subsection{Diffusion Entropy}

The Diffusion Entropy (DE) is defined as the Shannon entropy of the diffusion process $X(t)$ and was extensively used in the scaling detection of complex time series \cite{grigolini2001asymmetric,akin_jsmte09}.  
The DE algorithm is the following:
\begin{enumerate}
    \item Given a time lag $\Delta t$, split the time series $X(t)$ into overlapping time windows of duration $\Delta t$ and compute:
    $\Delta X (t, \Delta t) = X(t+\Delta t) - X(t),\ \forall\ t \in [0,t-\Delta t]$.
    \item For each time lag \(\Delta t\), evaluate the distribution \(p(\Delta x,\Delta t)\).
    \item Compute the Shannon entropy:
    \beq
    S(\Delta t) = - \int_{- \infty}^{+ \infty} p(\Delta x,\Delta t) \log p(\Delta x,\Delta t) dx
    \label{de_shannon}
    \eeq
    If the probability density function (PDF) is self-similar, i.e., \(p(\Delta x,\Delta t) = f(\Delta x / \Delta t^\delta)/\Delta t^\delta\), it results:
    \beq
    S(\Delta t) = A + \delta \log (\Delta t + T)
    \label{de_scaling}
    \eeq
    To check the validity of Eq. \eqref{de_scaling} on the data and evaluate the exponent $\delta$ is it sufficient to carry out a linear best fit with a logarithmic scale on the time axis.
\end{enumerate}

\section{\uppercase{Numerical simulations and results}}
\label{sec:results}

\begin{figure*}[hbt!]
\centering
\includegraphics[scale=0.43]{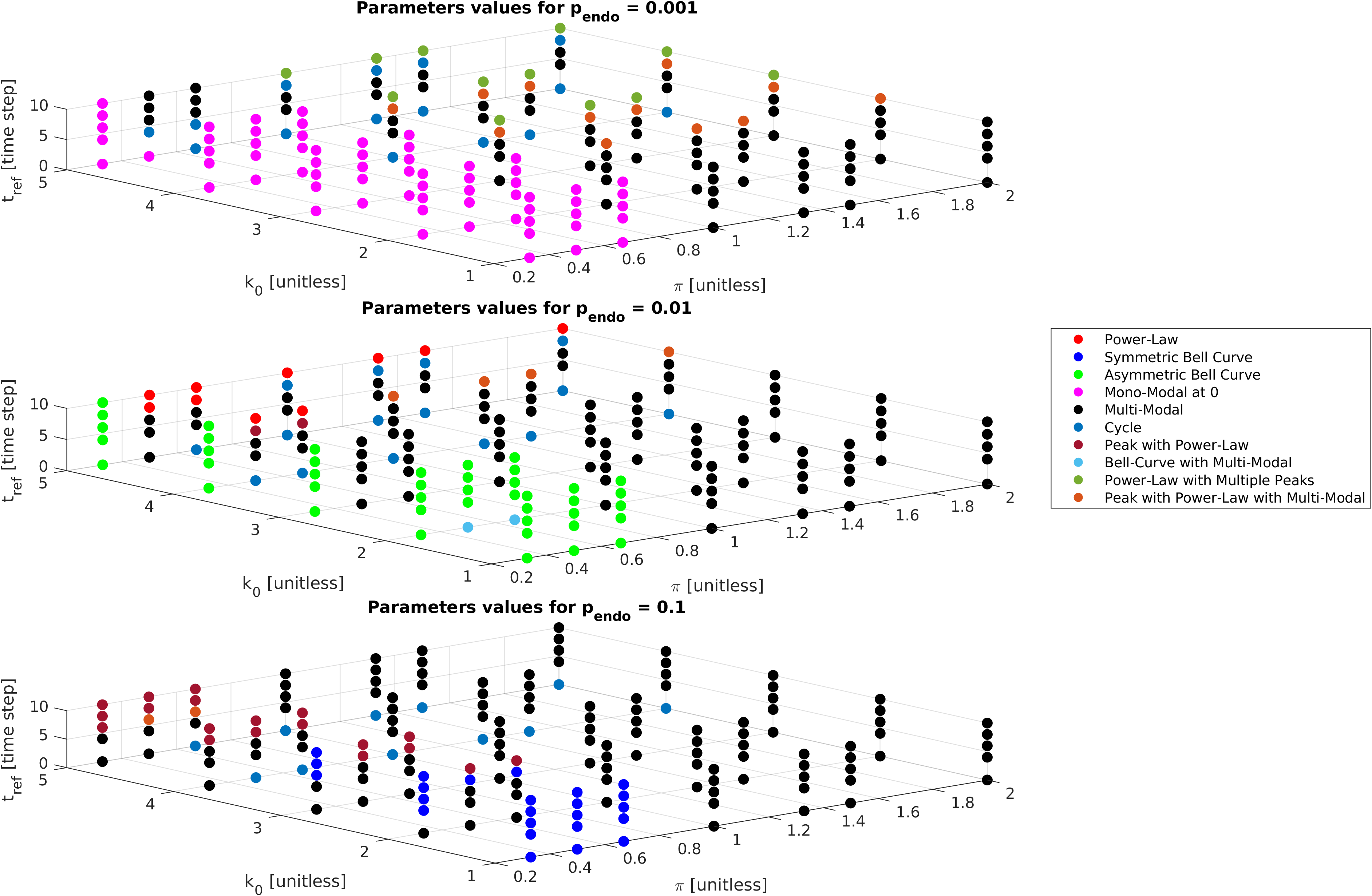}
\caption{Results of parameter analysis derived from the behaviour of the total activity distribution in ER networks.} 
\label{ER_parametric_analysis}
\end{figure*}

\begin{figure*}[hbt!]
\centering
\includegraphics[scale = 0.43]{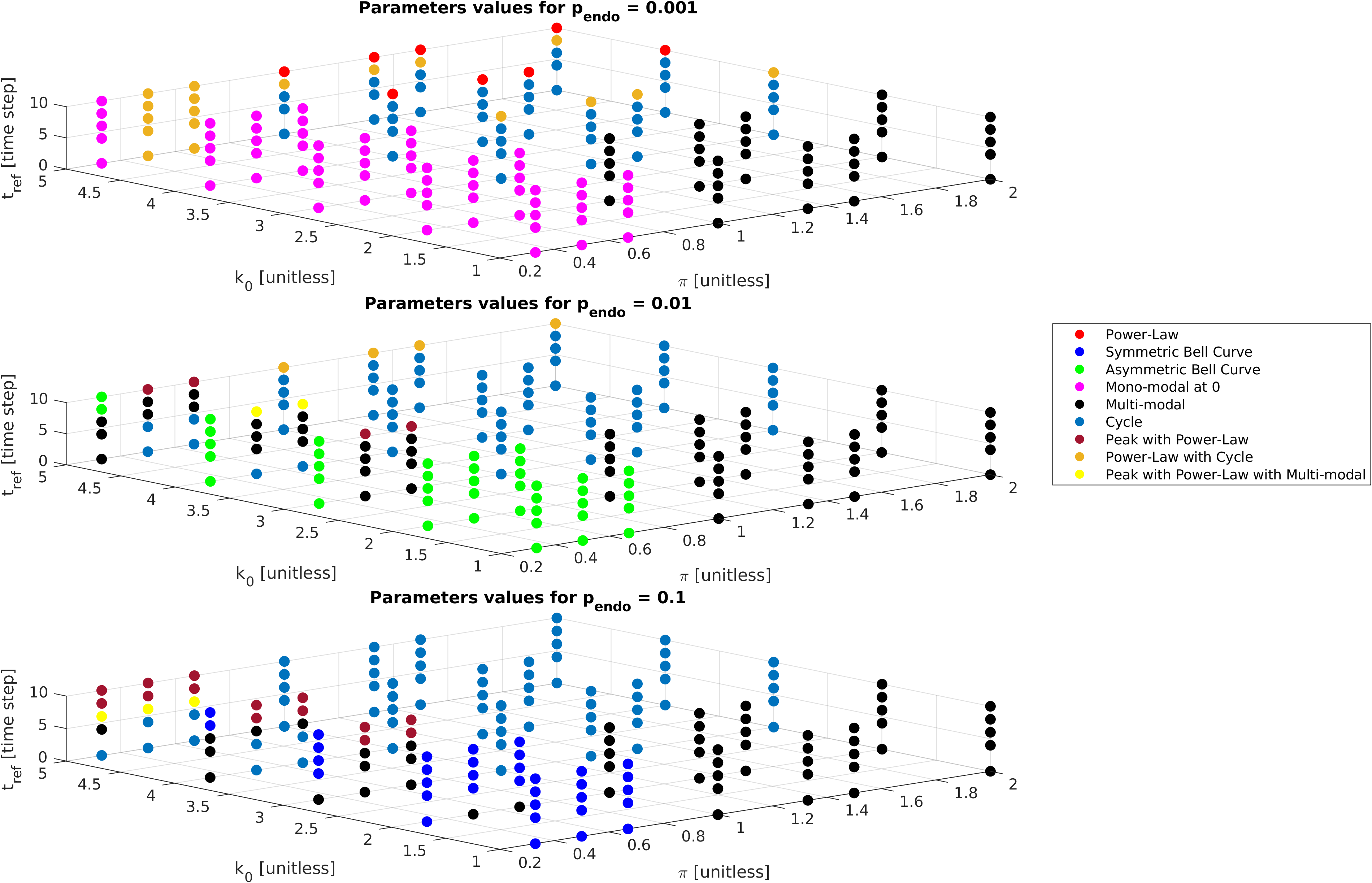}
\caption{Results of parameter analysis derived from the behaviour of the total activity distribution in SF network.} 
\label{SF_parametric_analysis}
\end{figure*}

\begin{figure*}[hbt!]
\centering
\subfloat[DFA]{\includegraphics[width=.51\textwidth]{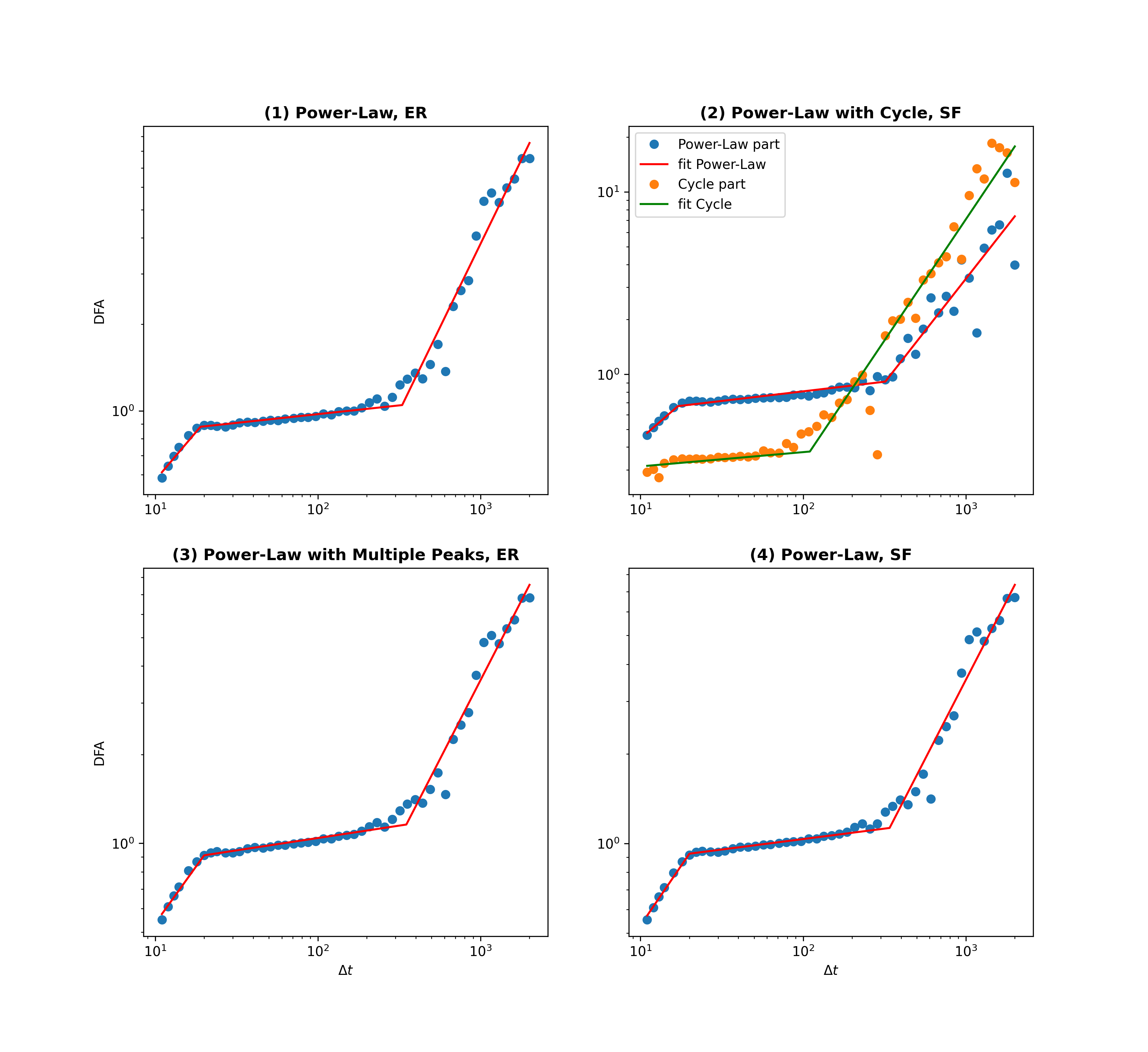}}
\label{fig:dfa_res}
\hspace{-.7cm}
\subfloat[DE]{\includegraphics[width=.51\textwidth]
{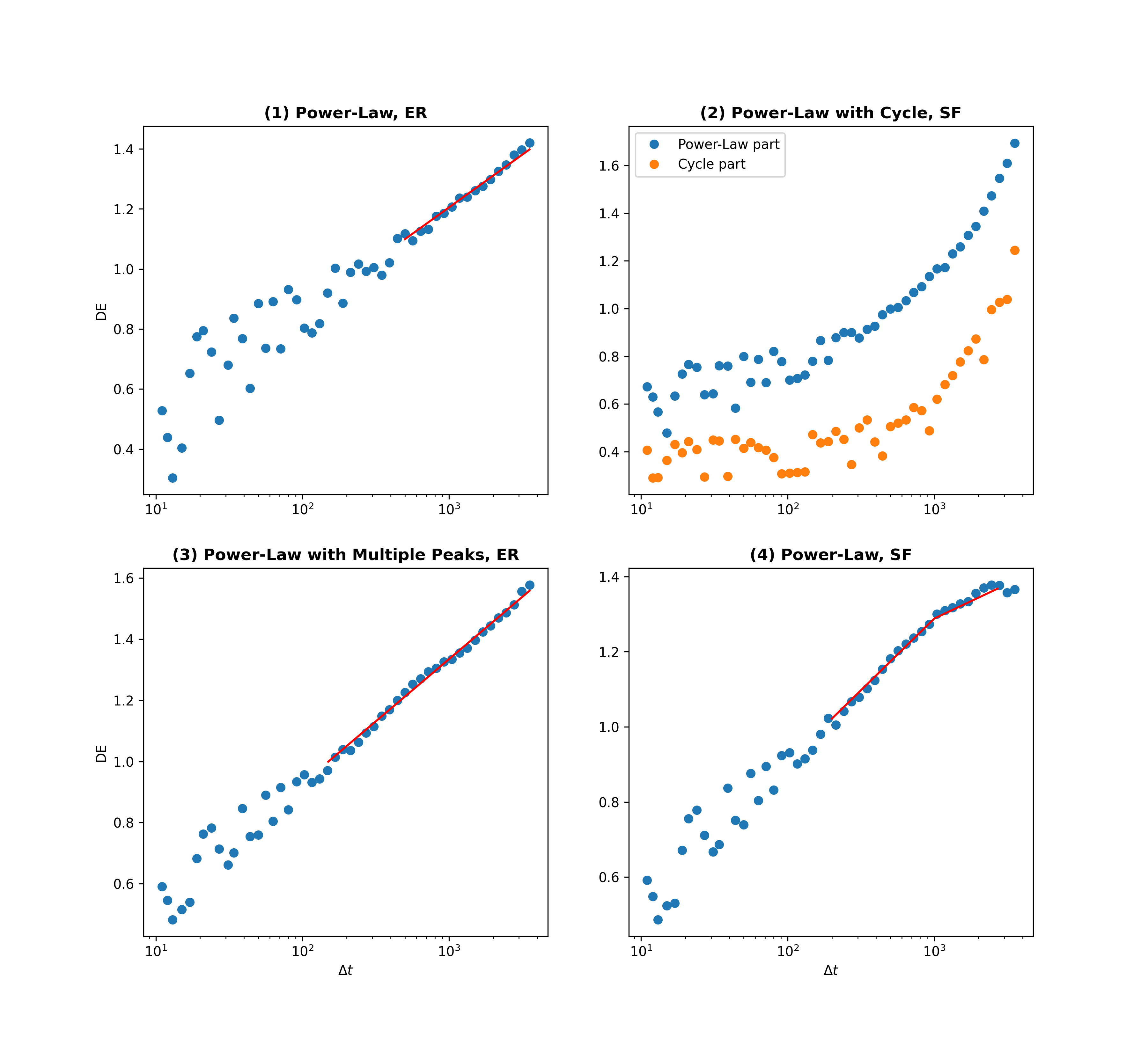}}
\label{fig:de_res}
\caption{DFA and DE analyses. Panels (1) and (2) same parameter set 
\(k_0\) = 5, \(t_{ref}\) = 10, \(b\) = 2, \(J\) = 3 and
\(p_{endo}\) = 0.01
(same as panels (9) of Fig. \ref{ER_average_activity_total_activity} and (9) of Fig. \ref{SF_average_activity_total_activity}). Panels (3) and (4) same as before but with different noise level \(p_{endo} = 0.001\)
(same as panels (10) of Fig. \ref{ER_average_activity_total_activity} and (8) of Fig. \ref{SF_average_activity_total_activity}).
}
\label{de_dfa}
\end{figure*}

\begin{figure*}[h!]
\centering
\subfloat[]{\includegraphics[scale=0.30]{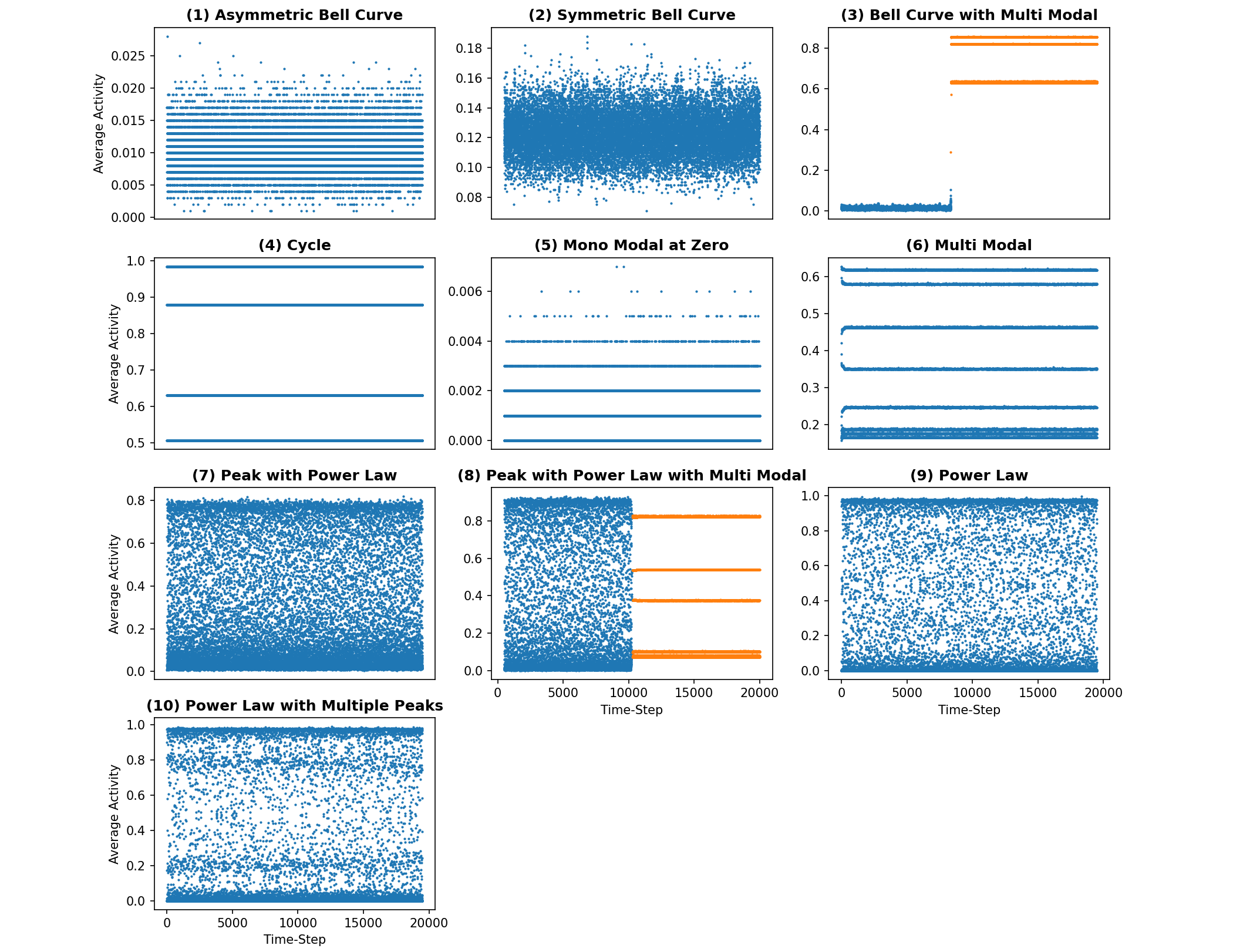}}
\hspace{-5.cm}
\subfloat[]{\includegraphics[scale=0.30]{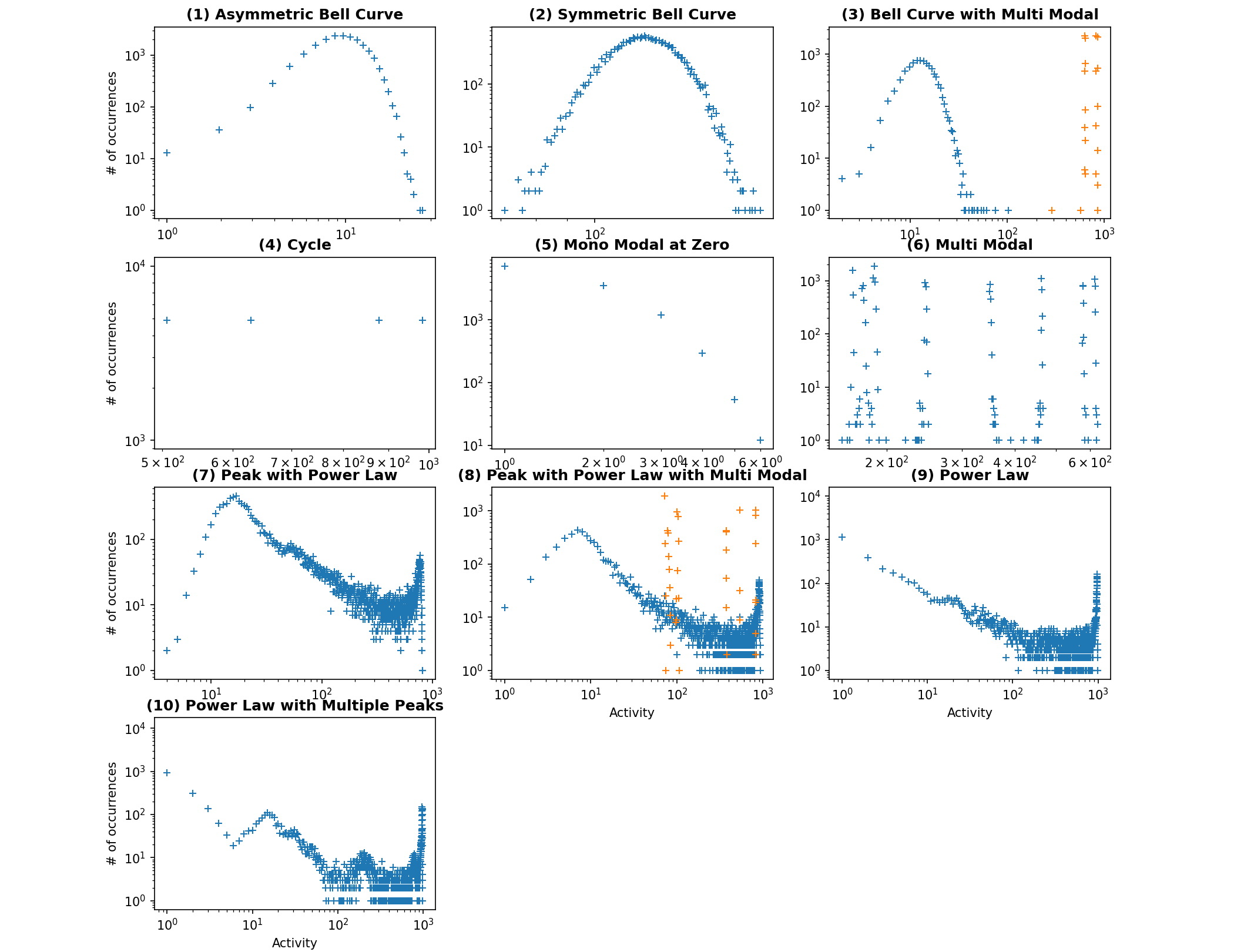}}
\caption{(a) Average Activity plots over time and (b) Histograms of Total Activity for all the qualitative behaviours found in ER networks. Parameters for each panel: (1) \(k_0\) = 1, \(t_{ref}\) = 4, \(b\) = 2, \(p_{endo}\) = 0.01, and \(J\) = 1; (2) \(k_0\) = 1, \(t_{ref}\) = 4, \(b\) = 2, \(p_{endo}\) = 0.1, and \(J\) = 1; (3) \(k_0\) = 2, \(t_{ref}\) = 0, \(b\) = 2, \(p_{endo}\) = 0.01, and \(J\) = 1; (4) \(k_0\) = 4, \(t_{ref}\) = 0, \(b\) = 2, \(p_{endo}\) = 0.01, and \(J\) = 1; (5) \(k_0\) = 1, \(t_{ref}\) = 4, \(b\) = 2, \(p_{endo}\) = 0.001, and \(J\) = 1; (6) \(k_0\) = 3, \(t_{ref}\) = 6, \(b\) = 3, \(p_{endo}\) = 0.1, and \(J\) = 2; (7) \(k_0\) = 5, \(t_{ref}\) = 6, \(b\) = 3, \(p_{endo}\) = 0.1, and \(J\) = 1; (8) \(k_0\) = 5, \(t_{ref}\) = 6, \(b\) = 3, \(p_{endo}\) = 0.1, and \(J\) = 2; (9) \(k_0\) = 5, \(t_{ref}\) = 10, \(b\) = 2, \(p_{endo}\) = 0.01, and \(J\) = 3; (10) \(k_0\) = 5, \(t_{ref}\) = 10, \(b\) = 2, \(p_{endo}\) = 0.001, and \(J\) = 3.}
\label{ER_average_activity_total_activity}
\end{figure*}

\begin{figure*}[h!]
\centering
\subfloat[]{\includegraphics[scale=0.32]{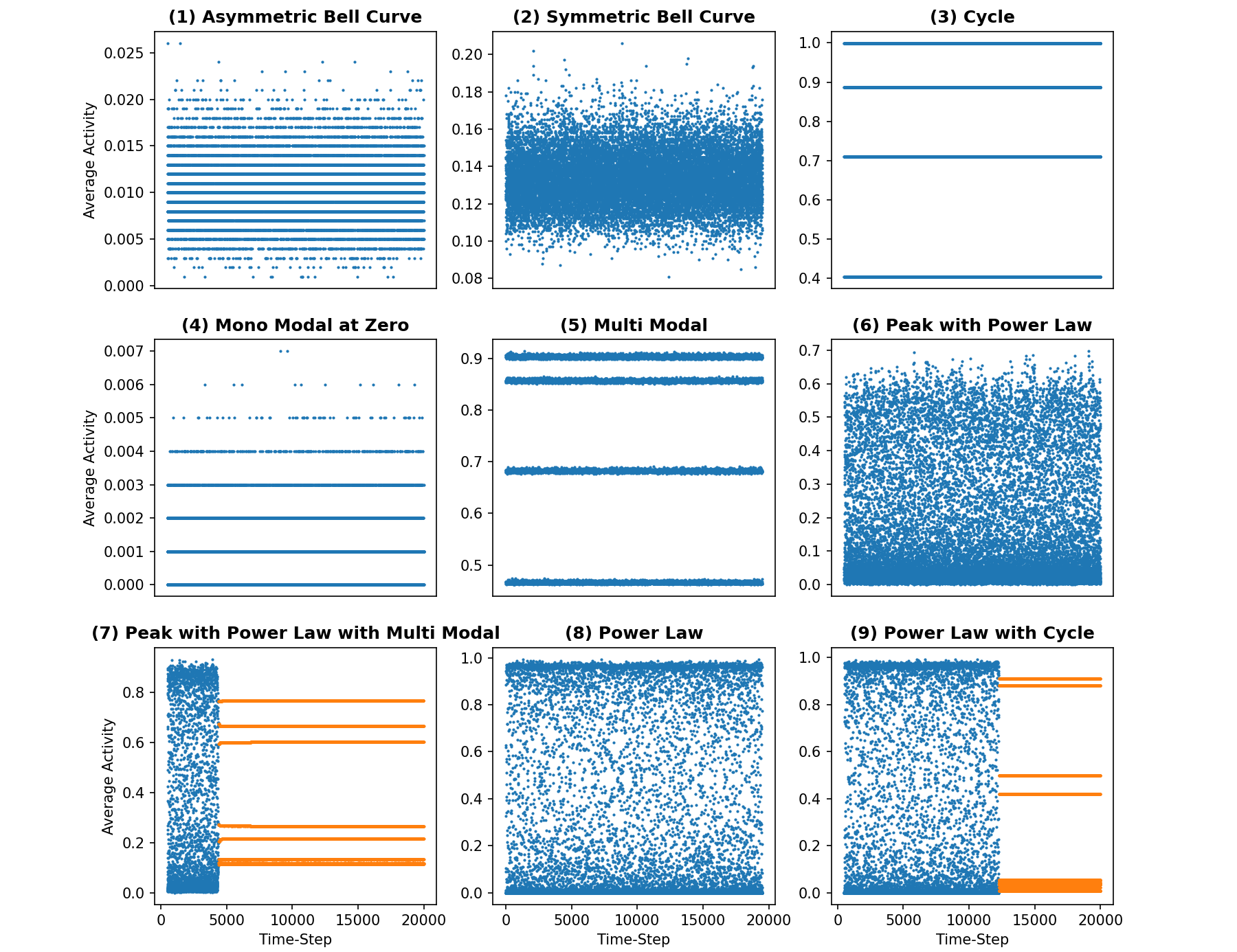}}
\hspace{5.cm}
\subfloat[]{\includegraphics[scale=0.32]{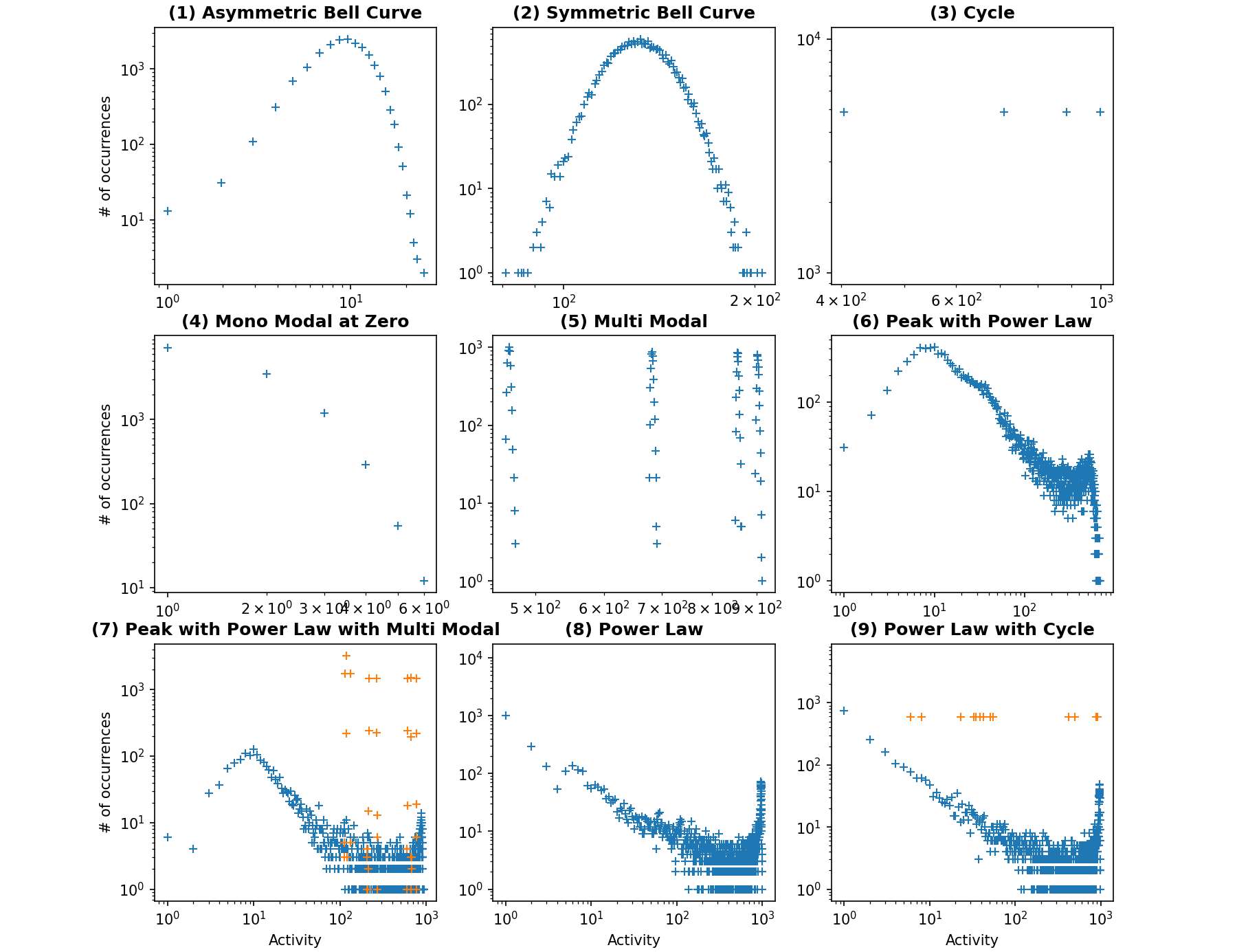}}
\caption{(a) Average Activity plots over time and (b) Histograms of Total Activity for all the qualitative behaviours found in SF networks. Parameters for each panel: (1) \(k_0\) = 1, \(t_{ref}\) = 0, \(b\) = 2, \(p_{endo}\) = 0.01, and \(J\) = 1; (2) \(k_0\) = 1, \(t_{ref}\) = 0, \(b\) = 2, \(p_{endo}\) = 0.1, and \(J\) = 1; (3) \(k_0\) = 5, \(t_{ref}\) = 0, \(b\) = 3, \(p_{endo}\) = 0.01, and \(J\) = 2; (4) \(k_0\) = 1, \(t_{ref}\) = 0, \(b\) = 2, \(p_{endo}\) = 0.001, and \(J\) = 1; (5) \(k_0\) = 3, \(t_{ref}\) = 0, \(b\) = 3, \(p_{endo}\) = 0.1, and \(J\) = 1; (6) \(k_0\) = 3, \(t_{ref}\) = 10, \(b\) = 3, \(p_{endo}\) = 0.1, and \(J\) = 2; (7) \(k_0\) = 5, \(t_{ref}\) = 10, \(b\) = 3, \(p_{endo}\) = 0.1, and \(J\) = 2; (8) \(k_0\) = 5, \(t_{ref}\) = 10, \(b\) = 2, \(p_{endo}\) = 0.001, and \(J\) = 3; (9) \(k_0\) = 5, \(t_{ref}\) = 10, \(b\) = 2, \(p_{endo}\) = 0.01, and \(J\) = 3.} \label{SF_average_activity_total_activity}
\end{figure*}

We performed a comprehensive parametric analysis with a fixed \(t_{max}\) value of 3, while systematically varying the other parameters. These include \(J \) which was constrained to integer values ranging from 1 to 4, \(p_{endo}\) set to $0.001$, $0.01$ and $0.1$, \(k_0\) limited to integers spanning from 1 to 5, \(b \) with two options of either 2 or 3, and \(t_{ref}\) set at 0, 4, 6, 8 and 10.
For dimensional reasons, the model's dynamics depend only on the
adimensional parameter \(\pi = \frac{J}{b}\), which is conveniently
used in the plots summarising the parametric analysis, i.e.
Figs. \ref{ER_parametric_analysis} and \ref{SF_parametric_analysis}.
The simulations were carried out for $20000$ time steps within networks comprising $N=1000$ neurons.
We analyzed the network behaviours by examining two key metrics: the total activity distribution $P(n_c|N_c = 1)$
and the average activity over time. 
The parametric analysis results are reported in Fig. \ref{ER_parametric_analysis} and Fig. \ref{SF_parametric_analysis}. 
For the ER networks we have identified the following qualitative behaviors in the total activity distributions:
\begin{enumerate}
    \item[1.] Asymmetric Bell Curve distribution
    \item[2.] Symmetric Bell Curve distribution
    \item[3.] Bell curve and transition to multi-modal distribution
    \item[4.] Cycle distribution
    \item[5.] Mono-modal at zero distribution
    \item[6.] Multi-modal distribution
    \item[7.] Peak with Power-Law distribution
    \item[8.] Peak with Power-Law and transition to multi-modal distribution
    \item[9.] Power-Law distribution
    \item[10.] Power-Law with Multiple Peaks distribution    
\end{enumerate}
For the SF networks we have identified the following qualitative behaviors in the total activity distributions:
\begin{enumerate}
    \item[1.] Asymmetric Bell Curve distribution
    \item[2.] Symmetric Bell Curve distribution
    \item[3.] Cycle distribution
    \item[4.] Mono-modal at zero distribution
    \item[5.] Multi-modal distribution
    \item[6.] Peak with Power-Law distribution
    \item[7.] Peak with Power-Law and transition to multi-modal distribution
    \item[8.] Power-Law distribution
    \item[9.] Power-Law and transition to cycle distribution   
\end{enumerate}
Figures \ref{ER_average_activity_total_activity} and \ref{SF_average_activity_total_activity} report the results for ER and SF networks, respectively. \\
We selected specific cases for further investigation using TC/IDC analysis. Specifically, we analyzed the WTs between successive coincidence events by applying the DFA and DE analyses. We report some of the results in Fig. \ref{de_dfa}, where
a few relevant cases involving power-law behaviour were selected.
The best-fit values of the scaling exponents $H$ and $\delta$ are 
reported in Table \ref{results_DFA_DE}.


%
%
%
%

\begin{table*}[hbt!]
        \caption{Result of fits for DFA and DE. For the ``Power-Law with Cycle (SF)'' row: Power-Law part is indicated with (PL), and Cycle part is indicated with (C). In the ``Power-Law (SF)'' case, the fit $\delta = 0.184$ for DE in the time range $\Delta t\sim 10^3-3\cdot 10^3$ was not reported.}
        \begin{center}
        \begin{tabular}{c|c|c|c|c|}
            \cline{2-5}
             & \multicolumn{2}{|c|}{DFA ($H$)} & \multicolumn{2}{|c|}{DE ($\delta$)}\\
            \cline{2-5}
             & Short-Time & Long-Time & Short-Time & Long-Time \\
            \hline
            \multicolumn{1}{|c|}{Power-Law (ER)} & 
            0.061 & 
            1.164 & / & 
            0.352 \\
            \hline
            \multicolumn{1}{|c|}{Power-Law (SF)} & 
            0.070 & 
            1.061 & / & 
            0.382 \\
            \hline
            \multicolumn{1}{|c|}{Power-Law with Multiple Peaks (ER)} &
            0.084 &  
            1.075 & / & 
            0.407 \\
            \hline
            \multicolumn{1}{|c|}{Power-Law with Cycle (SF)} &  
            0.105 (PL) 
            &  
            1.138 (PL) 
            & / & / \\
            \multicolumn{1}{|c|}{} & 
            0.078 (C)
            &  
            1.327 (C) 
            & / & /\\
            \hline
        \end{tabular}
        \end{center}
    \label{results_DFA_DE}
\end{table*}

%
%
%
%

\section{\uppercase{Discussion}}
\label{sec:discussion}

Our numerical simulations showed a large variety of behaviours in the Hopfield-type network in both network topologies, i.e., random (ER) and scale-free (SF). 
As can be seen from Figs. \ref{ER_average_activity_total_activity} and 
\ref{SF_average_activity_total_activity}, most qualitative behaviors
can be found in both network topologies, but with different sets of parameters.
However, some differences can be seen: (i) ``bell curve with multimodal'' and ``power-law with multiple peaks'' behaviours appear only in the ER network topology 
(panels (3) and (10) in Fig. \ref{ER_average_activity_total_activity}), while
(ii) ``power-law with cycle'' behaviour is exclusive of the SF network topology
(panels (9) of Fig. \ref{SF_average_activity_total_activity}). 
Interestingly, all other behaviours are seen in both topologies, even if with slight differences. In particular,
the emergence of power-law behaviour in the total activity distribution
(compare, e.g., panel (9) of Fig. \ref{ER_average_activity_total_activity} with
panel (8) of Fig. \ref{SF_average_activity_total_activity}).
It is worth noting that different kinds of power-law behaviours are seen in both topologies (panels (7-10) of
Fig. \ref{ER_average_activity_total_activity} and panels (6-9) of 
Fig. \ref{SF_average_activity_total_activity}). 
As can be seen also from Figs. 
\ref{ER_parametric_analysis} and \ref{SF_parametric_analysis}, for the pure power-law
behaviour (red dots) the main difference seems to lie in the different noise levels:
$p_{endo} = 0.01$ for ER and $p_{endo} = 0.001$ for SF.

\noindent
Another remarkable observation regards the abrupt transition among different
behaviours seen in some specific cases. In particular, some cases display a
initial power-law behaviour that can persist for a very long time, but
then it is followed by an unexpected transition to multimodal or cycle behaviour
(panels (8) in Fig. \ref{ER_average_activity_total_activity} and panels (7) and (9)
in Fig. \ref{SF_average_activity_total_activity}).
Only in ER networks, it is also seen a transition between a mono-modal to a multi-modal distribution where maxima are shifted towards
higher values of total activity (panels (3) of Fig. \ref{ER_average_activity_total_activity} ).

\noindent
Fig. \ref{de_dfa} shows some relevant behaviours in both DE and DFA functions.
In particular, we chose to analyze: (i) pure power-law behavior in both topologies
(panels (1) and (4)),
which surprisingly arises for similar parameter sets apart from the noise level;
(ii) power-law with multiple peaks in ER (panel (3)) and (iii) power-law with a transition to a cycle for SF (panel (2)).
Regarding ``power-law with cycle'' (panel (2)), due to the rapid transition from the power-law to the cycle behaviour, 
DFA and DE were applied separately to the two regimes.
Surprisingly, the cycle regimes gives a pattern of DFA and DE similar
to that of the power-law regime, even if with different slopes. 
Reliable fit values for $H$ and $\delta$ are reported in Table \ref{results_DFA_DE}. It can be seen that the qualitative behaviour
of DFA are essentially the same in the different cases. All the 
investigated cases have the same parameters, except for the noise level,
which is given by $p_{endo}=0.01$ for the top panels and $0.001$ for the bottom panels.
In summary, we have: (i) short-time with
very low $H$, associated with highly anti-persistent correlations; (ii) long-time with very high $H \sim 1$, except in panel (3) where $H>1$, associated with highly persistent correlations and superdiffusion.
Interestingly, we get $H \simeq 1$ for $p_{endo}=0.001$ in both 
topologies, while the pure power-law, which occurs for different noise
levels in the two topologies gives a larger value of $H$ for the ER network ($H\simeq 1.16$).
The DE displays a power-law only in the long-time regime that is, at variance with the DFA, in agreement with a subdiffusive behaviour.
This is not directly related to the persistence of correlations, but
directly to the shape of the diffusion PDF. In summary, in the long-time
regime of time lags, the diffusion generated by the coincidence events,
which are a manifestation of self-organizing behaviour, shows 
highly persistent correlations, as revealed by DFA, associated with a subdiffusive behaviour in the DE analysis.
This could be compatible with a very slow power-law decay in the WT-PDF,
i.e.,  $\psi(\tau) \sim \/\tau^\mu$ with $\mu<2$.

\section{\uppercase{Concluding remarks}}
\label{sec:conclusion}

Here we have investigated a Hopfield-type model and,
in particular, the model proposed in \cite{grinstein2005model},
being a simple prototype of bio-inspired neural
model. This includes bio-inspired features, such as the refractory time and the maximum firing time that can be tuned and interpreted, in the
context of AI, as hyper-parameters.
A particularly interesting bio-inspired feature is also encoded in the learning mechanism of Hopfield-type networks, which is exactly the Hebbian bio-inspired, unsupervised, learning.
The network topology is recognized to also play a central role in both global
network dynamics and learning efficiency.
This last feature is of particular interest in the AI field.

In particular, some authors focused on the effect of topological structure on learning features
\cite{kaviani_eswa2021}, in some cases finding
a better learning performance associated with
specific topologies, e.g., scale-free and/or small-world \cite{Lu_Chaos2023}.

\noindent
The present work represents a preliminary investigation of the relationships among connectivity features and temporal complexity of a simple spiking neural network without learning algorithms.
Interestingly, different topological structures
can give similar dynamical behaviours and complexity
features (see, e.g, Fig. \ref{ER_average_activity_total_activity}b, panel (9), with Fig. \ref{SF_average_activity_total_activity}b, panel (8), and Fig. \ref{de_dfa}a, panels (1) and (4)). 

\noindent
Regarding the relationship between connectivity and temporal complexity, further
investigations are needed to better understand the
reason why very different topologies can give similar complexity.
We expect these further investigations to deepen the
understanding of the relationship between dynamical
features of the network, e.g., temporal complexity,
connectivity structure and learning features, such as
storage capacity.

\noindent
We also plan future investigations regarding the relationship
between connectivity structure and learning algorithms, to study how the performance measures, jointly with the evaluation of complexity indices, change with the number of stored patterns (e.g., in the Hopfield model).




\section*{\uppercase{Funding}}
This work was supported by the Next-Generation-EU programme under the funding schemes PNRR-PE-AI scheme (M4C2, investment 1.3, line on AI)
FAIR “Future Artificial Intelligence Research”, grant id PE00000013, Spoke-8: Pervasive AI.

%
%


\bibliographystyle{apalike}
{\small
\bibliography{Bibliografia}}


\end{document}